\documentclass[twocolumn,showpacs,amsmath,amssymb,aps,prb,superscriptaddress,tbtags]{revtex4-2}

\usepackage[]{graphicx}
\usepackage[]{verbatim}
\usepackage[]{color}
\usepackage{overpic}
\usepackage{rotating}
\usepackage{mathtools}
\usepackage{appendix}
\usepackage{ulem}
\usepackage[margin=0.6in]{geometry}
\usepackage{tabularx}
\usepackage{float}
\usepackage{multirow}

\usepackage[]{hyperref}
 \hypersetup{
 colorlinks = true,
 linkcolor = blue,
 urlcolor = blue,
 citecolor = blue}
\usepackage{array}
\newcolumntype{L}[1]{>{\raggedright\let\newline\\\arraybackslash\hspace{0pt}}m{#1}}
\newcolumntype{C}[1]{>{\centering\let\newline\\\arraybackslash\hspace{0pt}}m{#1}}
\newcolumntype{R}[1]{>{\raggedleft\let\newline\\\arraybackslash\hspace{0pt}}m{#1}}
\newcolumntype{N}{@{}m{0pt}@{}}

\makeatletter
\newsavebox{\@brx}

\newcommand{\llangle}[1][]{\savebox{\@brx}{\(\m@th{#1\langle}\)}%
  \mathopen{\copy\@brx\mkern2mu\kern-0.8\wd\@brx\usebox{\@brx}}}
\newcommand{\rrangle}[1][]{\savebox{\@brx}{\(\m@th{#1\rangle}\)}%
  \mathclose{\copy\@brx\mkern2mu\kern-0.8\wd\@brx\usebox{\@brx}}}

  \newcommand{\lllangle}[1][]{\savebox{\@brx}{\(\m@th{#1\langle}\)}%
  \mathopen{\copy\@brx\copy\@brx\mkern4mu\kern-0.7\wd\@brx\usebox{\@brx}}}
\newcommand{\rrrangle}[1][]{\savebox{\@brx}{\(\m@th{#1\rangle}\)}%
  \mathclose{\copy\@brx\copy\@brx\mkern4mu\kern-0.7\wd\@brx\usebox{\@brx}}}

\graphicspath{{./}{./}}

\begin{document}
\title{Reconciling \texorpdfstring{$\pi$}{pi} phase shift in Josephson junction experiments with even-parity superconductivity in \texorpdfstring{Sr$_2$RuO$_4$}{Sr2RuO4}}
%Reconciling odd-parity evidence with even-parity superconductivity in \texorpdfstring{Sr$_2$RuO$_4$}{Sr2RuO4}}
\author{Austin W.~Lindquist}
\affiliation{Department of Physics and Center for Quantum Materials, University of Toronto, 60 St.~George St., Toronto, Ontario, M5S 1A7, Canada}
\author{Hae-Young Kee}
\email{hy.kee@utoronto.ca}
\affiliation{Department of Physics and Center for Quantum Materials, University of Toronto, 60 St.~George St., Toronto, Ontario, M5S 1A7, Canada}
\affiliation{Canadian Institute for Advanced Research, Toronto, Ontario, M5G 1Z8, Canada}
\begin{abstract}
The superconducting state of Sr$_2$RuO$_4$ was once thought to be a 
leading candidate for $p$-wave superconductivity. 
A constant Knight shift below the transition temperature provided
evidence for spin-triplet pairing, and a $\pi$ phase shift observed in 
Josephson junction tunneling experiments suggested odd-parity pairing,
both of which are described by $p$-wave states.
However, 
with recent experiments observing a significant decrease in the 
Knight shift below the transition temperature, signifying a 
spin-singlet state,
the odd-parity results are left to be reconciled.
In this work, we show that an even-parity 
pseudospin-singlet state originating from interorbital pairing via
spin-orbit coupling can
explain what has been assumed to be evidence for an odd-parity state.
In the presence of small mirror symmetry breaking,  
interorbital pairing is uniquely capable of displaying 
odd-parity characteristics required to explain these experimental results.
Further, we discuss how these experiments may be used to differentiate
the proposed pairing states 
of Sr$_2$RuO$_4$.
\end{abstract}
\maketitle

%\the\columnwidth

\section{Introduction}

The puzzle of superconductivity in Sr$_2$RuO$_4$ (SRO) has been
a longstanding problem with many seemingly contradictory 
experimental results 
\cite{Maeno1994Nature,Mackenzie2003RMP,Kallin2012RPP,Mackenzie2017NPJ}.
Once a leading candidate for $p+ip$-wave spin-triplet
superconductivity \cite{Rice1995JPCM,Ishida1998Nature}, 
recent experiments show a drop in the Knight
shift below the superconducting transition temperature, 
potentially ruling this
state out %, and possibly favoring a spin-singlet superconducting state
\cite{Pustogow2019Nature,Ishida2019,Chronister2021pnas}.
Other notable experiments suggest a two component order
parameter \cite{Okuda2002JPSJ,Benhabib2020, Ghosh2020} which breaks time-reversal
symmetry \cite{Luke1998Nature,Luke2000pbcm,Xia2006prl,Grinenko2020} 
and features gap nodes \cite{Deguchi2004jpsj,Hassinger2017PRX,Sharma2020}.
Attempts to explain these results have led to the recent proposals
of 
various multicomponent, even-parity pairing states 
\cite{Romer2019PRL,Suh2020PRR,Kivelson2020npj,Yuan2021PRB,Clepkens2021PRR,Clepkens2021PRB,Romer2021PRB,Romer2022PRR,Wang2022}.
%This collection of experimental data has led to the 
%recent proposals of a 
%a $d_{x^2-y^2}+ig_{xy(x^2-y^2)}$-wave state
%\cite{Kivelson2020npj,Yuan2021PRB,Clepkens2021PRB,Wang2022},
%an $s+id_{xy}$-wave state \cite{Clepkens2021PRR,Romer2021PRB}, 
%an $s+id_{x^2-y^2}$-wave state \cite{Romer2019PRL,Romer2022PRR},
%and a $d_{xz}+id_{yz}$-wave state \cite{Suh2020PRR}.

The proposed even-parity states are capable of explaining
many experimental results,
%While even-parity superconducting states can explain many
%experiments, 
however, little progress has been made in explaining the 
experimental data supporting
%results consistent with %which appear to be explained by %directly suggest 
odd-parity
superconductivity \cite{Nelson2004,Kidwingira2006s,Jang2011,Cai2020}.
Primarily, phase-sensitive
Josephson junction experiments observe a $\pi$ phase shift
of the superconducting order parameter under 
inversion \cite{Nelson2004}.
Previous studies have shown these results to be consistent with 
odd-parity pairing \cite{Asano2003PRB,Kawai2017PRB}, %however,
whereas 
conventional even-parity spin-singlet states
have remained in contradiction with these observations.
Conventional even-parity 
superconductors with inversion symmetry breaking have been shown to
display both even- and odd-parity character
in non-centrosymmetric superconductors
\cite{Hayashi2008physc,Klam2014PRB}, however, 
this effect would be much smaller in SRO. % than 
%noncentrosymmetric superconductors such as CsPt$_3$Si 
%\cite{Hayashi2008physc,Klam2014PRB}.

In this work, we study even-parity intraband pseudospin-singlet 
superconductivity, evolved from 
interorbital spin-triplet pairing via spin-orbit coupling (SOC)
in the presence of small mirror symmetry breaking hoppings
as a route to reconcile these remaining contradictions.
%We introduce the model used in this work, including 
The mirror symmetry 
breaking hopping term we introduce occurs
near surfaces, interfaces, or strain as sketched in the experimental 
setup shown in Fig.~\ref{fig1}a 
in Sec.~\ref{mh}.  
We then explain the setup of the Josephson junction 
calculations in Sec.~\ref{JJ} and show the current-phase relations for 
conventional even- and odd-parity pairing states.
Finally, in Sec.~\ref{iop} we present interorbital pairing and show that 
signatures of %the $d_{xy}$-, $g_{xy(x^2-y^2)}$-, and 
%$d_{xz}+id_{yz}$-wave states,
%all of which feature gap nodes in the tunneling direction,
pairing states which feature gap nodes in the tunneling direction 
may match those expected of an odd-parity state.  This behavior is 
made possible by the multiorbital nature of SRO and 
the intraband pseudospin-singlet pairing evolved from 
interorbital spin-triplet pairing.
%Finally, we consider how these states may provide the leading 
%contribution in the vicinity of surfaces or interfaces in Sec.~\ref{mf}.
%Since the proposed multicomponent states rely on a delicate
%accidental degeneracy, inversion symmetry breaking may disturb this
%balance and favor these states with the relevant nodal structure.

%as a route to reconcile these remaining contradictions.
%Through inversion symmetry breaking which occurs at surfaces or interfaces, 
%we find that the signatures of both the $d_{xy}$- and 
%$g_{xy(x^2-y^2)}$-wave superconducting states may match those 
%expected of an odd-parity pairing state.
%%we find that the signatures of this bulk superconducting state
%%can match that of an odd-parity pairing state.  
%The multiorbital
%nature of Sr$_2$RuO$_4$ is key in the origin of this behavior 
%and differentiates itself from conventional, intraorbital 
%superconductivity in these experiments.
%Further, we show that these states which feature nodes in the 
%tunneling direction may provide the leading contribution in the vicinity
%of the surface or interface, since the accidental degeneracy of the bulk
%relies on a delicate balance which is disturbed by inversion symmetry 
%breaking.
%%when 
%%the accidental degeneracy occuring in the bulk 
%%of the material is 
%%how these two pairing states may 

\section{Microscopic Hamiltonian}\label{mh}

\begin{figure}%[H]
\centering
\includegraphics{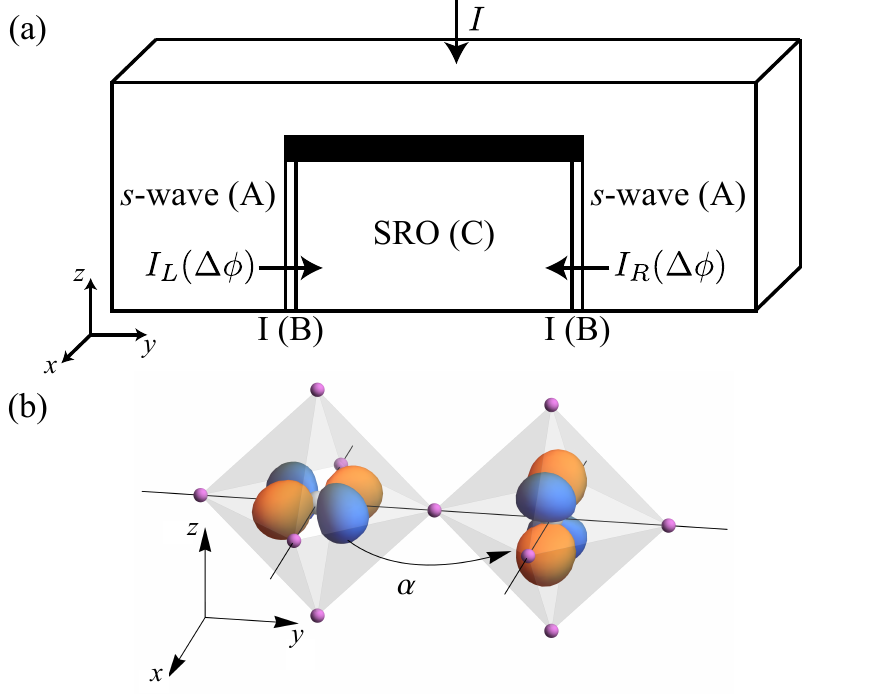}
\caption{
(a) Schematic of the Josephson junction setup showing %SRO in the middle region
%and interfaces to an $s$-wave superconductor on the left and right sides.
the $s$-wave superconductor, insulator (I), and SRO regions, denoted
$A$, $B$, and $C$ in the main-text equations, labeled
here in parentheses.
%The regions between SRO and the $s$-wave SC are modelled as
%two layers of a normal insulator.
In experiment SRO and the $s$-wave superconductor are separated in 
the $z$-direction by SiO, and no tunneling occurs at this interface,
signified by the filled black region here.
%In the vicinity of the interface where Josephson tunneling occurs, signified
%by the highlighted areas, inversion
%symmetry is broken, allowing for additional hopping terms.
(b) One example of interorbital hopping which is only allowed 
where mirror symmetry is broken.  When a $z$-mirror plane exists,
the orbital overlap is 0, but when broken, the overlap is finite.
}\label{fig1}
\end{figure}

The Josephson junction consists of a conventional single band 
$s$-wave superconducting
region, a normal insulator region, and the 
superconducting SRO region.
A schematic of the regions is shown in Fig.~\ref{fig1}(a), and they are
denoted $A$, $B$, and $C$, respectively, in the equations below.
Within regions $A$ and $B$ we use the single band kinetic Hamiltonian, 
written here for region $A$,
\begin{equation} 
%\begin{aligned}
H_{A} = \sum_{\textbf{k},i_y,\delta_y,\sigma}\xi_{A}(\textbf{k},\delta_y) 
c_{A,\textbf{k},i_y,\sigma}^{\dagger}c_{A,\textbf{k},i_y+\delta_y,\sigma}
%\\&+(\xi_A'(k_x)
%c_{A,i_y,k_x,\sigma}^{\dagger}c_{A,i_y+1,k_x,\sigma} + \text{h.c.}),
%\end{aligned}
\end{equation}
where $\xi_A(\textbf{k},\delta_y)$ is the electron dispersion in region $A$
between slabs at positions $i_y$ and $i_y+\delta_y$, and
%within a single slab, $xi_A'(k_x)$ is the dispersion featuring 
%nearest-neighbor interslab hopping, and 
$c_{A,\textbf{k},i_y,\sigma}^{\dagger}$ creates an electron in 
slab $i_y$ of region $A$ with momentum $\textbf{k}=(k_x,k_z)$ 
and spin $\sigma$. 
%We assume that the hopping between layers in the $z$-direction 
%is negligible in these regions.
The form of all dispersion terms as well as the values of the hopping 
parameters are given in Appendix \ref{tbs}.
%Note that the dispersion terms are written assuming translational 
%invariance in each region for convenience, however, calculations are performed without
%translational invariance in the $y$-direction, and so the real-space dispersion
%must be used in the $y$-direction.
The superconductivity in region $A$ is described by,
\begin{equation} 
H_A^\text{SC} = \sum_{\textbf{k},i_y}e^{i\phi_A}\Delta_A 
c_{A,\textbf{k},i_y,\uparrow}^{\dagger}c_{A,-\textbf{k},i_y,\downarrow}^{\dagger}+\text{h.c.,} 
\end{equation} 
where $\Delta_A$ is the $s$-wave order parameter, and $\phi_A$ is the 
superconducting phase.
Hopping between regions $A$ and $B$ is taken to have the same parameters
as hopping within either of the regions.

The normal state Hamiltonian of region $C$ is,
\begin{equation}
\begin{aligned}
&H_C = \sum_{\textbf{k},i_y,a,\sigma}\xi_C^a(\textbf{k},\delta_y)
c_{C,i_y,\textbf{k},\sigma}^{a\dagger}c_{C,i_y+\delta_y,\textbf{k},\sigma}^{a} \\
&+\sum_{\textbf{k},i_y,a\ne b,\sigma}\xi_C^{ab}(\textbf{k},\delta_y)
c_{C,\textbf{k},i_y,\sigma}^{a\dagger}c_{C,\textbf{k},i_y+\delta_y,\sigma}^{b}+\text{h.c.} + H_\text{SOC}, %\\
%&+i\lambda \sum_k \sum_{abl} \epsilon_{abl} c_{k,a,\sigma}^{3\dagger}c_{k,b,\sigma'}^3 \hat{\sigma}_{\sigma\sigma'}^l \\
%&+i\sum_{\textbf{k}\sigma\sigma'}\lambda_{\textbf{k}}\sigma_{\sigma\sigma'}^y
%c_{\textbf{k},xz,\sigma}^{3\dagger}c_{\textbf{k},xy,\sigma'}^{3} \\
%&-i\sum_{\textbf{k}\sigma\sigma'}\lambda_{\textbf{k}}\sigma_{\sigma\sigma'}^x
%c_{\textbf{k},yz,\sigma}^{3\dagger}c_{\textbf{k},xy,\sigma'}^{3},
\end{aligned}
\end{equation}
which includes intraorbital and interorbital dispersions, $\xi_C^a(\textbf{k},\delta_y)$ and 
$\xi_C^{ab}(\textbf{k},\delta_y)$, respectively, where $a$ and $b$ are the orbital
indices representing the $yz$, $xz$, and $xy$ orbitals,
as well as SOC terms.
%The precise form of the dispersions and the SOC are given in 
%Appendix \ref{tbs}.
%as well as atomic SOC, $\lambda$, 
%where $\epsilon_{abl}$ is the completely antisymettric tensor,
%and the momentum dependent $B_{2g}$ SOC, $\lambda_k$, 
%where $\lambda_k = 4\lambda_{B_{2g}}\sin k_x \sin k_y$.
Finally, the Hamiltonian describing hopping between regions $B$ and $C$,
where the interface occurs between $i_y = 1$ and $2$,
has the form,
\begin{equation}
H_\text{int} = \sum_{\textbf{k},a,\sigma} \xi_\text{int}^{a}(\textbf{k}) 
c_{C,\textbf{k},2,\sigma}^{a\dagger}c_{B,\textbf{k},1,\sigma}
%+t_\text{int}^{'a} c_{C,k_x,2,\sigma}^{a\dagger}c_{B,k_x\pm1,1,\sigma} 
+ \text{h.c.,}
\end{equation}
which features orbital dependence in the SRO region, as denoted by $\xi_\text{int}^a(\textbf{k})$.
%where $c_{C,i,j,\sigma}^{a\dagger}$ is the creation operator
%for $(x,y)$-coordinates $(i,j)$, in region $C$, and orbital $a$.
%The parameter $t_\text{int}^a$ $[t_\text{int}^{'a}]$ describes
%the [next] nearest-neighbor hopping between a site $(i,1)$ at the boundary of region $B$,
%and site $(i,2)$ $[(i\pm1,2)]$ at the boundary 
%of region $C$, with orbital index $a$.
%and the border between
%regions 2 and 3 is assumed to exist between $x$-coordinates 1 and 2.
%for layer $n$, with $y$-coordinate $j$ in region $m$.

We also consider the effects of mirror symmetry breaking in the 
$z$-direction (out of plane direction).  
This effect is largest near the surface normal to the $z$-direction,
but imperfections of the interface 
leading to a broken mirror plane
in the $z$-direction have previously been proposed to occur
\cite{Zutic2004PRL},
where it was shown that the experimental results may be 
explained by a $d_{xz}+id_{yz}$-wave state if the 
tunneling directions tilt out of the $xy$ plane.
Additionally, the growth of Au$_{0.5}$In$_{0.5}$ 
directly onto SRO to create the junction
may cause strain in SRO.  Any deformations of the
lattice that this leads to
may further
contribute to broken mirror symmetry throughout the sample.
%\textcolor{red}{
Dislocations may also contribute to this
mirror symmetry breaking, and have been found to 
occur near interfaces \cite{Ying2013nc}. %}
The lack of mirror symmetry 
in the $z$-direction means that hopping between the $xy$ and $xz(yz)$ 
orbitals is allowed to be finite in the $y(x)$ direction.
An example of such hopping is shown in Fig.~\ref{fig1}(b) for the $xy$ to $xz$ 
interorbital hopping.
These hoppings have the form:
\begin{equation}
\begin{aligned}
&h_\textbf{k}^{ISB}=
-\alpha[2i\sin k_x(c_{\textbf{k},i_y,\sigma}^{yz\dagger} c_{\textbf{k},i_y\sigma}^{xy} 
-c_{\textbf{k},i_y,\sigma}^{xy\dagger} c_{\textbf{k},i_y,\sigma}^{yz} ) \\
&- (c_{C,\textbf{k},i_y,\sigma}^{xz\dagger} c_{C,\textbf{k},i_y+1,\sigma}^{xy} 
-c_{C,\textbf{k},i_y,\sigma}^{xy\dagger} c_{C,\textbf{k},i_y+1,\sigma}^{xz} + \text{h.c.})],
\end{aligned}
\end{equation}
%In real space, these hoppings have the form:
%\begin{equation}
%\begin{aligned}
%\alpha &(c_{i,\sigma}^{xz\dagger} c_{i+y,\sigma}^{xy} -c_{i,\sigma}^{xy\dagger} c_{i+y,\sigma}^{xz} 
%+ c_{i,\sigma}^{yz\dagger} c_{i+x,\sigma}^{xy} -c_{i,\sigma}^{xy\dagger} c_{i+x,\sigma}^{yz} ),
%\end{aligned}
%\end{equation}
where $\alpha$ represents the hopping integral, which depends on the strength of
the mirror symmetry breaking, and
the use of $\delta_y=1$ here represents 
nearest-neighbor hopping between slabs.
%In momentum space this has the form,
%\begin{equation}
%\begin{aligned}
%h_k^{ISB}=&-2i\alpha[ \sin k_y(c_{k,\sigma}^{xz\dagger} c_{k,\sigma}^{xy} -c_{k,\sigma}^{xy\dagger} c_{k,\sigma}^{xz} ) \\
%&+\sin k_x(c_{k,\sigma}^{yz\dagger} c_{k,\sigma}^{xy} -c_{k,\sigma}^{xy\dagger} c_{k,\sigma}^{yz} )].
%\end{aligned}
%\end{equation}

In the next section, %we consider how the terms presented here affect the 
%CPRs for interorbital superconductivity.
we describe the setup for the Josephson tunneling calculations and apply
it with conventional $s$- and $p$-wave superconducting states in the 
SRO region.  
Then, we consider interorbital superconductivity and show how
the current-phase relation (CPR) is affected by the mirror symmetry breaking terms 
introduced here, showing that they may behave like 
the conventional $s$-wave state, or potentially more like 
the $p$-wave state, depending on
the nodal structure as well as the strength of the mirror
symmetry breaking.

\section{Josephson Calculations}\label{JJ}

To calculate the Josephson CPR, we 
use the lattice Green's function method presented in 
Ref.~\onlinecite{Kawai2017PRB},
which considers only $p$-wave pairing to explain
experimental results.
Semi-infinite Green's functions are obtained for the $s$-wave
and SRO regions using the 
recursive Green's function approach \cite{Umerski1997PRB}.
A single layer of the normal insulator is added
on the surfaces of both of these regions by the Dyson equations,
\begin{equation}
\hat{G}_0^B(\textbf{k},i\omega_l) = (i\omega_l-\hat{u}_0(\textbf{k})-\hat{t}_{0,-1}
\hat{G}_{-1}^A(\textbf{k},i\omega_l)\hat{t}_{-1,0})^{-1},
\end{equation}
\begin{equation}
\hat{G}_1^B(\textbf{k},i\omega_l) = (i\omega_l-\hat{u}_1(\textbf{k})-\hat{t}_{1,2}
\hat{G}_{2}^C(\textbf{k},i\omega_l)\hat{t}_{2,1})^{-1},
\end{equation}
where the interface is in the $xz$ plane.
Here, $\hat{G}_n^m(\textbf{k},i\omega_l)$ is the Green's function of layer
$n$ in region $m$, $\hat{u}_n(\textbf{k})$ is the part 
of the Hamiltonian of layer $n$, and $\hat{t}_{n,n+1}$ is the 
part of the Hamiltonian featuring hopping the the $y$-direction,
%nonlocal 
%part of the Hamiltonian, 
representing hopping between layers $n$ and 
$n+1$.
The left and right systems are combined using the two equations,
\begin{equation}
\hat{G}_{00}(\textbf{k},i\omega_l)=\{[\hat{G}_0^B(\textbf{k},i\omega_l)]^{-1}-
\hat{t}_{01}\hat{G}_1^B(\textbf{k},i\omega_l)\hat{t}_{10}\}^{-1},
\end{equation}
\begin{equation}
\hat{G}_{11}(\textbf{k},i\omega_l)=\{[\hat{G}_1^B(\textbf{k},i\omega_l)]^{-1}-
\hat{t}_{10}\hat{G}_0^B(\textbf{k},i\omega_l)\hat{t}_{01}\}^{-1}.
\end{equation}
These are then used to obtain the nonlocal Green's functions
\begin{equation}
\hat{G}_{01}(\textbf{k},i\omega_l) = \hat{G}_0^B(\textbf{k},i\omega_l)\hat{t}_{01}
\hat{G}_{11}(\textbf{k},i\omega_l),
\end{equation}
\begin{equation}
\hat{G}_{10}(\textbf{k},i\omega_l) = \hat{G}_1^B(\textbf{k},i\omega_l)\hat{t}_{10}
\hat{G}_{00}(\textbf{k},i\omega_l).
\end{equation}
From this, the CPR, $I(\Delta\phi)$, is obtained,
\begin{equation}
\begin{aligned}
I(\Delta\phi)&=\frac{iet}{\hbar}\int_{-\pi}^{\pi}\text{Tr}'\frac{1}{\beta}\sum_l 
[\hat{G}_{01}(\textbf{k},i\omega_l,\Delta\phi) \\ &-\hat{G}_{10}(\textbf{k},i\omega_l,\Delta\phi)]d\textbf{k},
\end{aligned}
\end{equation}
where Tr$'$ represents a trace over only the electron space,
$\beta=\frac{1}{k_BT}$, $t$ is the nearest-neighbor hopping integral in 
the normal region, and $\Delta\phi=\phi_A-\phi_C$ represents the phase
difference in the superconducting phase of the $s$-wave and SRO regions.

\begin{figure}%[H]
\includegraphics{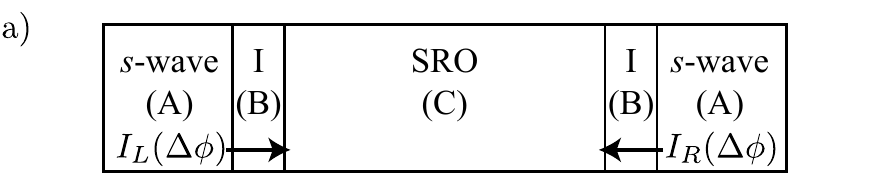}
\includegraphics{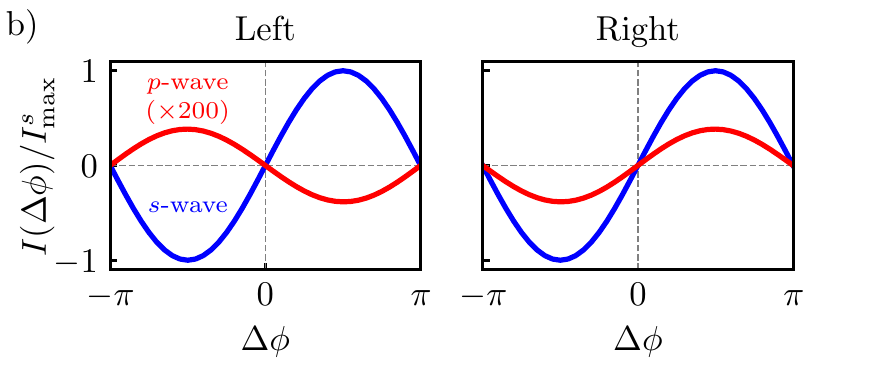}
\caption{
(a) Schematic of the setup for CPR calculations, showing the left and right
interfaces between the $s$-wave SC and SRO.
(b) Current-phase relations for both  $s$-wave SC/I/SRO interfaces
with the SRO region modelled using an intraorbital $s$- and $p$-wave
pairing state.
}\label{swave}
\end{figure}

%To understand how conventional superconductors behave within 
%a Josephson junction, we first consider $s$- and $p$-wave
%states within the SRO region.

%We first consider the CPRs of conventional superconducting states.
To understand the CPRs of the Josephson $\pi$ junction, we first review
conventional superconducting states.
First, we use an intraorbital $s$-wave state in the SRO region,
\begin{equation}
H_{C,s}^\text{SC} = \sum_{\textbf{k},i_y,a}e^{i\phi_C}\Delta_C^a 
c_{C,\textbf{k},i_y,\uparrow}^{a\dagger}c_{C,-\textbf{k},i_y,\downarrow}^{a\dagger}+\text{h.c.,} 
\end{equation} 
where the order parameter, $\Delta_C^a$, exists within all three orbitals.
The explicitly written superconducting phase, $\phi_C$, is kept the same 
between all three orbitals, and calculations were performed with
$\Delta_C^{yz}=\Delta_C^{xz}=-\Delta_C^{xy}$, similar to 
intraorbital $s$-wave contributions found in Ref.~\onlinecite{Puetter2012EPL}.
%(1) $\Delta_3^a$ having the same sign in all orbitals and 
%(2) $\Delta_3^{xy}$ having the opposite sign of $\Delta_3^{yz}=\Delta_3^{xz}$.
Fig.~\ref{swave}(b) shows the CPR at the left and right interfaces
of the $\pi$ junction for the $s$-wave pairing state in blue.
This shows no phase shift between the two 
interfaces, as expected for an even-parity pairing state.

Next, we use an intraorbital $p_x+ip_y$-wave pairing state,
\begin{equation}
\begin{aligned}
&H_{C,p}^\text{SC} = \sum_{\textbf{k},i_y,a}e^{i\phi_C}d_{z}^a 
(2\sin k_x c_{C,\textbf{k},i_y,\uparrow}^{a\dagger}
c_{C,-\textbf{k},i_y,\downarrow}^{a\dagger} \\ 
&+c_{C,\textbf{k},i_y,\uparrow}^{a\dagger}
c_{C,-\textbf{k},i_y+1,\downarrow}^{a\dagger} 
-c_{C,\textbf{k},i_y,\uparrow}^{a\dagger}
c_{C,-\textbf{k},i_y-1,\downarrow}^{a\dagger}) 
+\text{h.c.} 
\end{aligned}
\end{equation} 
Again, the phase is kept the same between all three orbitals, and 
the same sign convention is used as in the $s$-wave pairing state.
Fig.~\ref{swave}(b) shows the CPR at the left and right interfaces 
for the $p$-wave pairing state in red. 
Now, the $\pi$ phase shift is observed between the interfaces, 
characteristic of the odd-parity state.

These two results show a clear distinction between 
even- and odd-parity pairing. 
Since the current measured through the entire junction in experiment
depends on the addition of these two curves, the odd-parity phase difference 
corresponds to a minimum 
current through the junction, as measured in Ref.~\onlinecite{Nelson2004},
whereas the even-parity state would correspond to a maximum.
While this allows for 
straightforward differentiation between intraorbital pairing states,
in the next section we consider how this changes when
interorbital pairing is instead considered, in the presence
of broken mirror symmetry.

\section{Interorbital Superconductivity}\label{iop}

\begin{figure}%[H]
\includegraphics{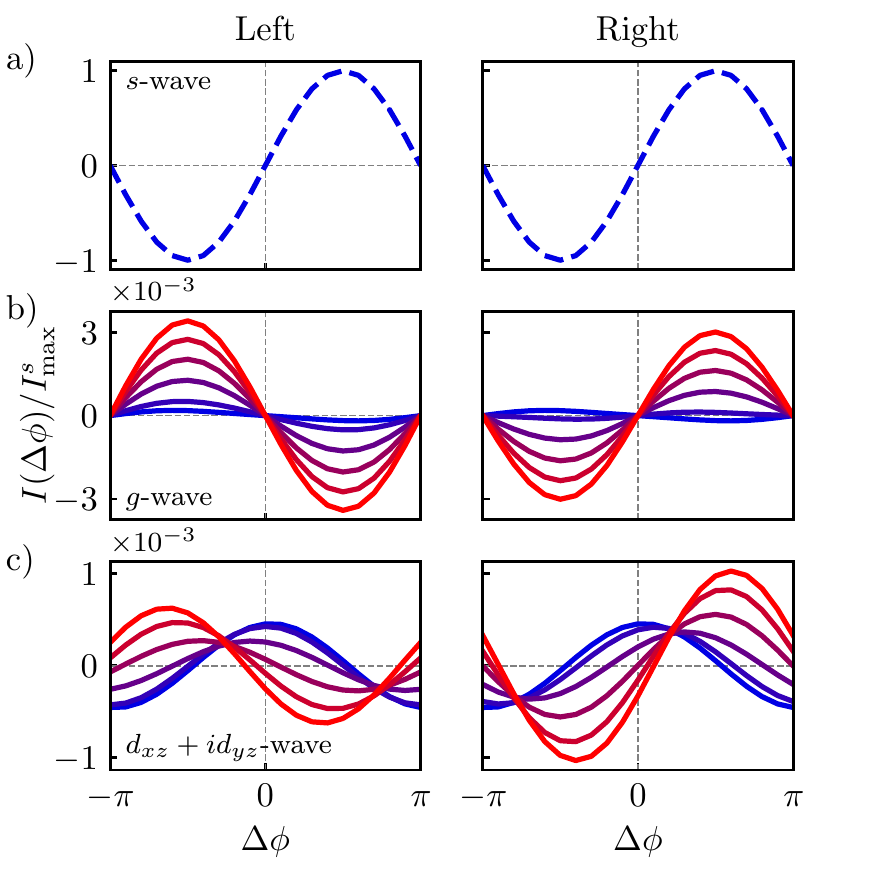}
\caption{Current-phase relations for an interorbital SC/I/$s$-wave interface with 
a varying mirror symmetry breaking hopping, increasing linearly from 
0 to 0.005 in units of $2t=1$ between the blue and red curves.
The left column represents the left interface, and the right
represents the right interface.  The SRO region is considered here with {\it interorbital} 
(a) $s$- %(solid) and $d_{x^2-y^2}$- (dashed), 
%(c) $d_{xy}$-, and 
(b) $g_{xy(x^2-y^2)}$-,
and (c) $d_{xz}+id_{yz}$-wave pairing.
Only the $\alpha=0$ curve is shown with the dashed line in (a) as no significant 
change is observed in the presence of $\alpha$.
The $d_{x^2-y^2}$- and $d_{xy}$-wave CPRs (not shown) are qualitatively similar to the $s$- 
and $g_{xy(x^2-y^2)}$-wave curves, respectively.
}\label{iosddg}
\end{figure}

Equipped with the techniques, we now perform the Josephson 
calculations in multiorbital superconductors.
In multiorbital systems, the orbital degree of freedom allows 
for additional types of pairing which satisfy the antisymmetric fermion
wavefunction requirement.  In addition to even-parity spin-singlet 
and odd-parity spin-triplet pairings, even-parity spin-triplet
and odd-parity spin-singlet pairings are possible when the wavefunction
is antisymmetric with respect to the orbital index.
The possibility of interorbital pairing occuring in SRO
has been considered in previous studies 
\cite{Puetter2012EPL,Hoshino2015PRL,Hoshino2016PRB,Ramires2016PRB,Cheung2019PRB,
Ramires2019PRB,Huang2019PRB,Gingras2019PRL,Kaba2019PRB,
Lindquist2020PRR,Lindquist2022PRR,Gingras2022PRB},
and, importantly, has recently been discussed as a microscopic route to 
the proposed  $s+id_{xy}$-wave \cite{Clepkens2021PRR}, 
$d_{x^2-y^2}+ig_{xy(x^2-y^2)}$-wave \cite{Clepkens2021PRB,Wang2022}, and 
$d_{xz}+id_{yz}$-wave pairing states \cite{Suh2020PRR}.

The general form of the interorbital superconducting state is written,
\begin{equation}\label{Delta}
\begin{aligned}
\hat{\Delta}_{a/b}^{l\dagger}=\frac{1}{4N}\sum_{\textbf{k},i_y}[i\hat{\sigma}^y \hat{\sigma}^l]_{\sigma \sigma'}
(c_{\textbf{k},i_y,\sigma}^{a\dagger} c_{-\textbf{k},i_y,\sigma'}^{b\dagger} \\ 
-c_{\textbf{k},i_y,\sigma}^{b\dagger} c_{-\textbf{k},i_y,\sigma'}^{a\dagger}).
\end{aligned}
\end{equation}
Here, the interorbital pairing %order parameter, $\Delta_{a/b}^l$, 
is a constant 
in $k$-space, i.e., $s$-wave.  However, momentum dependence may be 
revealed after transforming into the band basis due to momentum-dependent
SOC or a combination of dispersion terms and SOC
\cite{Clepkens2021PRR,Clepkens2021PRB}.

To understand the momentum dependence of the intraband gap that arises
in the presence of broken mirror symmetry, let us first consider 
a bulk two orbital model which features SOC and a mirror 
symmetry breaking hopping term.
We use the basis $\Psi_\textbf{k}^\dagger = (\psi_\textbf{k}^\dagger,\mathcal{T}\psi_\textbf{k}^T\mathcal{T}^{-1})$
where $\mathcal{T}$ represents time-reversal, and
$\psi_\textbf{k}^\dagger=(c_{\textbf{k},\uparrow}^{xz\dagger},c_{\textbf{k},\downarrow}^{xz\dagger},
c_{\textbf{k},\uparrow}^{xy\dagger},c_{\textbf{k}k,\downarrow}^{xy\dagger})$, 
The kinetic and SOC parts of the Hamiltonian are,
\begin{equation}\label{2orb}
\begin{aligned}
H_\textbf{k} = &~ \frac{1}{2}\xi^+(\textbf{k}) \rho_3\tau_0\sigma_0 + \frac{1}{2}\xi^-(\textbf{k}) \rho_3\tau_3 \sigma_0
+\lambda(\textbf{k}) \rho_3\tau_2\sigma_3 \\ &+ 
t(\textbf{k}) \rho_3\tau_1\sigma_0+\alpha(\textbf{k}) \rho_3\tau_2\sigma_0,
\end{aligned}
\end{equation}
where $\rho$, $\tau$, and $\sigma$ are Pauli matrices representing 
the particle-hole, orbital, and spin bases, respectively. 
The orbital dispersions are 
$\xi^\pm(\textbf{k})=\xi^{xz}(\textbf{k})\pm\xi^{xy}(\textbf{k})$, the SOC is $\lambda(\textbf{k})$, 
the orbital hybridization is $t(\textbf{k})$, and mirror symmetry
breaking hopping $\alpha(\textbf{k})$. The precise momentum dependence of each
of these terms is left unspecified for this analysis, but all are given 
in Appendix \ref{tbs}
for the numerical calculations performed below.  Importantly, the mirror 
symmetry breaking hopping is an odd function in $\textbf{k}$, 
$\alpha(-\textbf{k})=-\alpha(\textbf{k})$, %where $\textbf{k}=(k_x,k_y)$,
while all other terms are even functions of the form $f(-\textbf{k})=f(\textbf{k})$.

The interorbital-singlet 
spin-triplet pairing with a $d$-vector in the $z$-direction is written as 
%\begin{equation}
$H_\text{SC}=\Delta_z \rho_1 \tau_2 \sigma_3$. % where
%$\Delta_z = \langle\hat\Delta_{xz/xy}^z\rangle$.
%\end{equation}
Transforming $H_\text{SC}$ to the band basis, i.e., the
basis in which $H_\textbf{k}$ is diagonal, the intraband pseudospin-singlet 
and pseudospin-triplet pairing %in one of the
%bands (labeled $\beta$)
can both be identified,
%is found,
\begin{equation}
\begin{aligned}
H_\text{SC}&=\Delta^S(\textbf{k})
(\hat\Delta_{\textbf{k},0}^\alpha-\hat\Delta_{\textbf{k},0}^\beta)
%(c_{-\textbf{k},\downarrow}^\alpha c_{\textbf{k},\uparrow}^\alpha
%- c_{-\textbf{k},\uparrow}^\alpha c_{\textbf{k},\downarrow}^\alpha
%-c_{-\textbf{k},\downarrow}^\beta c_{\textbf{k},\uparrow}^\beta \\
%&+ c_{-\textbf{k},\uparrow}^\beta c_{\textbf{k},\downarrow}^\beta)
+ \Delta^T(\textbf{k})
(\hat\Delta_{\textbf{k},z}^\alpha-\hat\Delta_{\textbf{k},z}^\beta),
%(c_{-\textbf{k},\downarrow}^\alpha c_{\textbf{k},\uparrow}^\alpha
%+ c_{-\textbf{k},\uparrow}^\alpha c_{\textbf{k},\downarrow}^\alpha \\ 
%&-c_{-\textbf{k},\downarrow}^\beta c_{\textbf{k},\uparrow}^\beta
%- c_{-\textbf{k},\uparrow}^\beta c_{\textbf{k},\downarrow}^\beta).
\end{aligned}
\end{equation}
where the pairing operators are written 
$\hat\Delta_{0(z)}^\alpha(\textbf{k})=c_{-\textbf{k},\downarrow}^\alpha c_{\textbf{k},\uparrow}^\alpha
\mp c_{-\textbf{k},\uparrow}^\alpha c_{\textbf{k},\downarrow}^\alpha$. 
The pseudospin-singlet and -triplet coefficients are given by,
\begin{equation}\label{sgap}
\begin{aligned}
\Delta^S(\textbf{k})&=\frac{-i\Delta_z\lambda(\textbf{k})}{\sqrt{\xi^-(\textbf{k})^2+4t(\textbf{k})^2+4\lambda(\textbf{k})^2}}
%\\ &+\frac{-4i\Delta_z\lambda(\textbf{k})\alpha(\textbf{k})^2}{[\xi^-(\textbf{k})^2+4t(\textbf{k})^2+4\lambda(\textbf{k})^2]^{3/2}}
+\cdots,
\end{aligned}
\end{equation}
and
\begin{equation}\label{tgap}
\begin{aligned}
\Delta^T(\textbf{k})&=\frac{i\Delta_z\alpha(\textbf{k})}{\sqrt{\xi^-(\textbf{k})^2+4t(\textbf{k})^2+4\lambda(\textbf{k})^2}}
%\\ &+\frac{4i\Delta_z\alpha(\textbf{k})\lambda(\textbf{k})^2}{[\xi^-(\textbf{k})^2+4t(\textbf{k})^2+4\lambda(\textbf{k})^2]^{3/2}}
+\cdots,
\end{aligned}
\end{equation}
where $(\cdots)$ represents higher order terms in the Taylor expansions.
Further details of these equations are given in Appendix \ref{2orbap}.
%The important result from these equations is that there exists a 
Importantly, there exists a 
pseudospin-singlet gap contribution proportional to $\lambda(\textbf{k})\Delta_z$, 
as well as a pseudospin-triplet contribution proportional to 
$\alpha(\textbf{k})\Delta_z$, both of the same order.
Note that %the since $\Delta_z$ is a constant, 
the momentum dependence of these
two gap coefficients are independent, and locations of nodes in the 
singlet gap do not necessarily correspond to nodes in the triplet gap.

%gap is found,
%\begin{equation}\label{gap}
%|\Delta_\text{band}^\pm|=\frac{i\Delta_z(\lambda_k\pm\alpha_k)}{\sqrt{(\xi_k^-)^2+(\lambda_k\pm\alpha_k)^2}}.
%\end{equation}
%This shows how the standard pseudospin-singlet gap forms at the Fermi level due to the SOC, and
%how it obtains its momentum dependence,
%the gap is proportional to $\lambda_k\Delta_z$, where $\lambda_k$ provides the momentum 
%dependent form factor.
%%This also displays how the pairing
%%at the Fermi level can have strong momentum dependence, through the momentum dependence
%%of $\lambda_k$.  
%This equation also shows that there is a second contribution to 
%the gap at the Fermi level from the inversion symmetry breaking hopping, proportional
%to $\alpha_k\Delta_z$.  This contribution is odd-parity pseudospin-triplet pairing.
%Importantly, since the interorbital order parameter is a constant, the
%momentum dependence of these two contributions to the gap are completely independent,
%and the location of nodes in the primary gap, $\lambda_k\Delta_z$, do not need to be 
%locations of nodes in the secondary gap, $\alpha_k\Delta_z$.
%This is unique from pairing induced from an intraorbital pairing state, 
%where nodes of the original intraorbital state must also be nodes 
%of the induced state.
%%This is contrary to any other types of pairing which may be induced from an intraorbital
%%pairing state, where the momentum dependence of the induced state will be directly
%%affected by the momentum dependence of the original state.

For comparison, 
this same analysis is applied to an intraorbital SC state of the 
form,
\begin{equation}
H'_\text{SC}=\Delta^+(\textbf{k})\rho_1\tau_0\sigma_0+\Delta^-(\textbf{k})\rho_1\tau_3\sigma_0,
\end{equation}
where $\Delta^\pm(\textbf{k}) = \frac{1}{2}(\Delta^{xz}(\textbf{k})\pm\Delta^{xy}(\textbf{k}))$ in this
two orbital model.  Again 
transforming this into the band basis, %the intraband gap terms are found to be
the pairing Hamiltonian is rewritten in terms of the intraband contributions,
\begin{equation}
\begin{aligned}
H'_\text{SC}=\Delta^{S_1'}&(\textbf{k})(\hat\Delta_0^\alpha+\hat\Delta_0^\beta)
+\Delta^{S_2'}(\textbf{k})(\hat\Delta_0^\alpha-\hat\Delta_0^\beta) \\
&+\Delta^{T'}(\textbf{k})(\hat\Delta_z^\alpha-\hat\Delta_z^\beta).
\end{aligned}
\end{equation}
The first singlet coefficient remains unchanged between the bases,
$\Delta^{S_1'}(\textbf{k})=\Delta^+(\textbf{k})$.  The other
singlet coefficient is,
\begin{equation}\label{intras}
\begin{aligned}
%\langle c_{k,\uparrow}^\beta c_{-k,\downarrow}^\beta&-
%c_{k,\downarrow}^\beta c_{-k,\uparrow}^\beta\rangle =
%\Delta^+(k) \\ 
\Delta^{S_2'}(\textbf{k})=\frac{\Delta^-(\textbf{k})\xi^-(\textbf{k})}{\sqrt{\xi^-(\textbf{k})^2+4t(\textbf{k})^2+4\lambda(\textbf{k})^2}}
+\cdots,
\end{aligned}
\end{equation}
and the triplet coefficient,
\begin{equation}\label{intrat}
\begin{aligned}
%\langle c_{k,\uparrow}^\beta c_{-k,\downarrow}^\beta&+
%c_{k,\downarrow}^\beta c_{-k,\uparrow}^\beta\rangle =
\Delta^{T'}(\textbf{k})=\frac{4\Delta^-(\textbf{k})\alpha(\textbf{k})\xi^-(\textbf{k})\lambda(\textbf{k})}
{[\xi^-(\textbf{k})^2+4t(\textbf{k})^2+4\lambda(\textbf{k})^2]^{3/2}}
+\cdots.
\end{aligned}
\end{equation}
%\begin{equation}
%\begin{aligned}
%|\Delta_\text{band}^{1\pm}|&=\Delta^++\frac{\Delta^-\xi_k^-}
%{\sqrt{(\xi_k^-)^2+4(t_k^2+(\lambda\pm\alpha)^2)}}, \\ 
%|\Delta_\text{band}^{2\pm}|&=\Delta^+-\frac{\Delta^-\xi_k^-}
%{\sqrt{(\xi_k^-)^2+4(t_k^2+(\lambda\pm\alpha)^2)}}.
%\end{aligned}
%\end{equation}
In this case, the odd-parity contribution is very small since it only 
shows up at a higher order expansion.  Additionally, since the momentum
dependence of the gap is encoded in $\Delta^-(\textbf{k})$, any nodes that appear within
the pseudospin-singlet pairing gap must also be nodes in the pseudospin-triplet
gap.  Therefore, the effect of the mirror symmetry breaking hopping term on
an intraorbital pairing state is
not expected to be significant for small $\alpha(\textbf{k})$. 
%In this case, there is no direct odd-parity contribution to the intraband gap.
%An odd-parity contribution still exists due to the inversion symmetry breaking, 
%however, the effect of this is considerably smaller than that from a direct 
%odd-parity term.
The contrast between these two cases is important.  For 
an interorbital SC state, it is possible to induce a non-zero odd-parity
gap where the original even-parity gap has a node.  However, for an 
intraorbital SC state, the induced odd-parity gap must have nodes wherever
the original even-parity state has nodes.

Based on Eqs.~(\ref{sgap}) and (\ref{tgap}), there are three significant contributions to the CPR
at a given interface that
we can write qualitatively,
\begin{equation}
\begin{aligned}
I(\phi) \sim 
c_1(\Delta \lambda(\textbf{k}))&\sin\phi + c_2(\Delta \alpha(\textbf{k}))\sin2\phi \\
&+ c_3(\Delta \alpha(\textbf{k})) \sin\phi,
\end{aligned}
\end{equation}
where $c_i$ are the coefficients of the various contributions.
The first, $c_1(\lambda(\textbf{k})\Delta_z)$, describes Josephson tunneling from the 
pseudospin-singlet state of SRO to the spin-singlet state of the $s$-wave 
superconductor.  The contributing gap is even parity and therefore this term
does not feature a $\pi$ phase shift between opposite interfaces.
Next, $c_2(\alpha(\textbf{k})\Delta_z)$ describes the first pseudospin-triplet to spin-singlet 
tunneling process.  Since this is direct pseudospin-triplet to spin-singlet 
tunneling, this term features a $\sin2\phi$ dependence.  %As was the case
%with the $p$-wave state above, this term is significant when the intermediate
%region is metallic.  
If the intermediate region is insulating,
then the $\sin 2\phi$ contribution is small, and in the presence of SOC, there 
is again a $\sin\phi$ with coefficient $c_3(\alpha(\textbf{k})\Delta_z)$ term that contributes. This term
is due to the odd-parity gap and therefore features a $\pi$ phase shift
between opposite interfaces.

The overall behaviour depends on the relative size of these coefficients.
The largest coefficient is generally $c_1(\lambda(\textbf{k})\Delta_z)$ since
$|\lambda(\textbf{k})|>|\alpha(\textbf{k})|$ is expected,
but when $\lambda(\textbf{k})$ features nodes, this is not always true.  For the 
$d_{xy}$- and $g_{xy(x^2-y^2)}$-wave states, nodes exist along the $x$-
and $y$-directions, so an interface in the $(010)$ direction may feature 
competition between the $\lambda(\textbf{k})\Delta_z$ gap and the $\alpha(\textbf{k})\Delta_z$ 
due to a small $\lambda(\textbf{k})$.
Alternatively, the $d_{xz}+id_{yz}$-wave state is gapless in the $z=0$ plane
and therefore any tunneling direction perpendicular to the $c$-axis would 
feature this competition.
In both cases, if the intermediate region is insulating, the %$\sin2\phi$ contribution
$c_2(\alpha(\textbf{k})\Delta_z)$ contribution
is small compared to the $c_3(\alpha(\textbf{k})\Delta_z)$ %$\sin\phi$ contribution, 
allowing for a path to
a $\pi$ phase shift from one of these even-parity states.

To confirm the above analysis works for the 3-orbital model of SRO,
calculations were performed on the full 3-orbital system using
various superconducting states.  Fig.~\ref{iosddg}(a) shows the CPR %for various 
%strengths of inversion symmetry breaking hopping 
for the $s$-wave state.  
The addition of the mirror symmetry breaking hopping has no significant 
effect on the CPR and is omitted in these plots.
This is consistent with expectations from the two-orbital analysis above, since 
%changes due to the inversion symmetry breaking term are
%small since 
the pairing state does not feature
nodes along the $(010)$ direction.
Fig.~\ref{iosddg}(b) shows the CPR 
for the $g_{xy(x^2-y^2)}$-wave
pairing state.  The effect of $\alpha$ is much more significant here,
and a $\pi$ phase shift is observed for a large enough 
value of $\alpha$. 
%In both cases the amplitude of the CPR changes sign for $\alpha$ around
%20\% of $\lambda_{B_{2g}}$.  However, these calculations were performed with a
%large $\lambda_{B_{2g}}$ and using a more realistic value of $\lambda_{B_{2g}}=$ 1.6 meV
%this sign change occurs with $\alpha \approx$ 10\%, or near 0.2 meV.
Fig.~\ref{iosddg}(c) uses the $d_{xz}+id_{yz}$-wave
pairing state where again the effect of $\alpha$ is significant.
In this case, the $\pi$ phase shift does not occur via a simple sign change
of the amplitude, but rather as a gradual phase shift due to competition between
the contributions from the $d_{xz}$ and $d_{yz}$ components. %Note that this means
%that the phase shift from this state does not need to be exactly 0 or $\pi$ like the
%single component states, but can take any value.
This shows that a $\pi$ phase shift is possible even when the pairing is
even-parity, as long as nodes are present in the gap structure and 
a small mirror symmetry breaking hopping is allowed.  

The $d_{x^2-y^2}$- and $d_{xy}$-wave states are not plotted, but appear very similar to
the $s$- and $g_{xy(x^2-y^2)}$-wave states, respectively. This is expected based on the 
nodal structure of each of these states in the $(010)$ direction. Note that if 
the tunneling direction were instead in the $(110)$ direction then behavior of 
the $d_{x^2-y^2}$- and $d_{xy}$-wave states is expected to switch, with 
$d_{x^2-y^2}$ instead showing the $\pi$ phase shift. In such a case, the 
combined $d_{x^2-y^2}+ig_{xy(x^2-y^2)}$-wave state is expected to exhibit a 
$\pi$ phase shift, similar to the $d_{xz}+id_{yz}$-wave state. 

\section{Discussion and summary}

A full understanding of the superconductivity of SRO
has remained elusive despite vast interest, as well as
a wide array of experiments 
performed on the material.
Recent results have pushed the community 
towards potentially adopting an even-parity spin-singlet pairing state,
although conventional states of this nature are not
able to consistently explain all observations.
Interorbital superconductivity has recently been shown
to provide a microscopic route to various multi-component pairing
states, each capable of explaining %many of the experimental results
a subset of the experimental results
\cite{Suh2020PRR, Clepkens2021PRR, Clepkens2021PRB, Wang2022}.
However, Josephson junction experiments 
capable of detecting the phase change of a superconducting 
order parameter between interfaces observe a $\pi$ phase
shift at opposite interfaces of SRO \cite{Nelson2004}.
While this is expected in odd-parity 
superconductors, these even-parity proposals are left to
provide some explanation for this observed behavior.

We've shown that using even-parity interorbital pairing, it
is possible to observe a $\pi$ phase shift between opposite 
interfaces in these Josephson junction experiments.
%$d_{x^2-y^2}+ig_{xy(x^2-y^2)}$-wave pairing state \cite{Clepkens2021PRB},
%which has been considered as a top candidate to explain many 
%experimental results \cite{Kivelson2020npj}.
%Interorbital pairing may also explain the Josephson 
%tunneling experiments, once thought to be smoking-gun
%evidence of an odd-parity pairing state.
The $s$-wave nature of the interorbital pairing in the orbital 
basis provides a route to finite odd-parity pseudospin-triplet
pairing in the band basis, induced by mirror symmetry breaking.
Importantly, the induced odd-parity pairing can have a nodal structure
independent of the even-parity state since the even- and odd-parity 
intraband 
pairings 
arise from interorbital
pairing via different orbital mixing terms.
%even where the even-parity pseudospin-singlet pairing features
%nodes.
Due to this, Josephson tunneling current in the direction of 
the even-parity nodes is dominated by the induced odd-parity 
behavior.
This is in contrast to the intraorbital pairing states, where
any induced odd-parity character features nodes in the same positions
as the original even-parity state, meaning the odd-parity character
does not dominate. 
While the precise tunneling direction within the $xy$-plane was
not determined in experiment \cite{Nelson2004}, the $d_{xz}+id_{yz}$-wave state
is completely gapless within the plane, and therefore these results are 
expected to hold for any tunneling direction.
The other proposals discussed here feature more complex gap structures and
therefore would be more sensitive to the tunneling direction.
If the multicomponent nature of the SC state is due to an accidental degeneracy, 
the precise gap structure where tunneling occurs may also be influenced 
significantly by surface effects.

Experimental determination of the directional dependence of the phase shift
would differentiate the proposed pairing states and could provide 
crucial information for identifying the superconducting pairing of 
SRO.
Specifically, %if no $\pi$ phase shift is observed in directions 
we expect to see no $\pi$ phase shift when tunneling occurs in a
direction 
far from the in-plane nodes of both gap components.
We therefore propose these phase sensitive Josephson
junction experiments with a known orientation of the
$a$ and $b$ axes as a route to differentiate the various 
pairing symmetry proposals.
%While leading proposals for the superconducting state do not feature
%nodes in the (100) or (010) directions, both the $s+id_{xy}$-
%and $d_{x^2-y^2}+ig_{xy(x^2-y^2)}$-wave states have one component
%with nodes in those directions.  
%If the accidental degeneracy is lifted near the surface or interfaces,
%these states may dominate the behavior and would therefore lead to the
%observed $\pi$ phase shift.

Although only one type of mirror symmetry breaking hopping is considered
here, others may contribute to this effect depending on the precise
geometry of the surface.
Here we focus on the phase shift observed in the Josephson
junction experiment \cite{Nelson2004}, however, experiments observing 
half-quantum 
vortices also provide evidence for odd-parity spin-triplet pairing, but 
use samples with a different geometry
\cite{Jang2011,Cai2020}.
These results may also occur due to interorbital pairing with
mirror symmetry breaking, but more work is needed to determine 
if these results are compatible with the pairing states discussed here.
%if the 
%effects are strong enough, and if so, which states are compatible with
%the results.
Additionally, spatially varying strain in a system with $d_{x^2-y^2}$-
and $g_{xy(x^2-y^2)}$-wave degeneracy has been proposed as being 
able to explain these half-quantum vortices \cite{Yuan2021PRB}.
While work still remains to distinguish the various multicomponent
even-parity pairing proposals, we believe that directionally 
resolved phase sensitive measurements can provide an important 
piece of puzzle of superconductivity in SRO.

\begin{acknowledgments} 
We acknowledge support from the
Natural Sciences and Engineering 
Research Council of Canada Discovery Grant 2022-04601. 
H.Y.K.~also acknowledges support from the Canadian Institute for 
Advanced Research and the Canada Research Chairs Program.
Computations were performed on the Niagara supercomputer
at the SciNet HPC Consortium. SciNet is funded by: the
Canada Foundation for Innovation under the auspices of
Compute Canada; the Government of Ontario; Ontario
Research Fund - Research Excellence; and the University
of Toronto.
\end{acknowledgments} 

\bibliography{SC-SDW}

\appendix

\section{Tight-Binding Model}\label{tbs}

The form of the microscopic Hamiltonian is introduced 
in Sec.~\ref{mh}.  Here, we provide the dispersion
terms used within that model, as well as the values used
for various hopping parameters throughout this work.
The dispersion of the $s$-wave and normal insulator 
regions are given by,
\begin{equation}
\begin{aligned}
\xi_{A(B)}(\textbf{k},0) &= - \mu_{A(B)}
-2 t\cos k_x , \\
\xi_{A(B)}(\textbf{k},1) &= t -2t'\cos k_x.
\end{aligned}
\end{equation}
All hopping terms are limited to next nearest neighbor, such
that the slab setup for the surface Green's functions
features one layer of atoms at a time.

Next, the dispersion terms in the SRO
region are given by,
\begin{equation}
\begin{aligned}
\xi_C^{yz(xz)}(\textbf{k},0)&=-\mu_\text{1D} - 2t_{2(1)}\cos k_{x}, \\
%\xi_C^{xz}(k_x,0)&=-\mu_{1D} -2t_1\cos k_{x} \\
\xi_C^{xy}(\textbf{k},0) &= 
-\mu_{xy} - 2t_{3}\cos k_{x}, \\ 
%\\&+8t_{x}^{xy}\cos \frac{k_x}{2}\cos\frac{k_y}{2}\cos\frac{k_z}{2}, \\
\xi_C^{yz(xz)}(\textbf{k},\frac{1}{2})&=4t_{z}^\text{1D}\cos \frac{k_x}{2}\cos\frac{k_z}{2}, \\
%\xi_C^{xz}(k_x,\frac{1}{2})&=4t_{z}^{1D}\cos \frac{k_x}{2}\cos\frac{k_z}{2}, \\
\xi_C^{xy}(\textbf{k},\frac{1}{2})&=4t_{z}^{xy}\cos \frac{k_x}{2}\cos\frac{k_z}{2}, \\
\xi_C^{yz(xz)}(\textbf{k},1)&= -t_{1(2)}, \\
\xi_{C}^{xy}(\textbf{k},1) &= -t_3 - 2t_{4}\cos k_x, \\
\xi_C^{yz/xz}(\textbf{k},\frac{1}{2})&= 4t_6i\sin \frac{k_x}{2} \cos \frac{k_z}{2}, \\
\xi_C^{yz/xy}(\textbf{k},\frac{1}{2})&= -4t_7\sin \frac{k_x}{2} \sin \frac{k_z}{2}, \\
\xi_C^{xz/xy}(\textbf{k},\frac{1}{2})&= 4t_7i\cos \frac{k_x}{2} \sin \frac{k_z}{2}, \\
\xi_C^{yz/xz}(\textbf{k},1)&= 2t_\text{1D}i\sin k_x.
\end{aligned}
\end{equation}
Here, $\delta_y=\frac{1}{2}$ occurs with interlayer hopping in the $z$-direction
due to the offset between layers.
%\xi_3^{yz(xz)}(\textbf{k})&=-\mu_{1D} -2t_1\cos k_{y(x)} - 2t_2\cos k_{x(y)} \\
%&+8t_{x}^{1D}\cos \frac{k_x}{2}\cos\frac{k_y}{2}\cos\frac{k_z}{2}, \\
%%\end{aligned}
%%\end{equation}
%%\begin{equation}
%%\begin{aligned}
%\xi_3^{xy}(\textbf{k}) &= 
%-\mu_{xy} - 2t_{3}(\cos k_{x} + \cos k_{y}) \\
%&+ 4t_{4}\cos k_x \cos k_y 
%\\&+8t_{x}^{xy}\cos \frac{k_x}{2}\cos\frac{k_y}{2}\cos\frac{k_z}{2}, \\
%%\end{aligned}
%%\end{equation}
%%\begin{equation}
%%\begin{aligned}
%\xi_3^{yz/xz}(\textbf{k}) &= -4t_\text{1D}\sin k_x \sin k_y
%\\&-8t_6\sin\frac{k_x}{2}\sin\frac{k_y}{2}\cos\frac{k_z}{2},\\
%%\end{aligned}
%%\end{equation}
%%\begin{equation}
%\xi_3^{yz(xz)/xy}(\textbf{k}) &= 
%-8t_7\sin\frac{k_{x(y)}}{2}\cos\frac{k_{y(x)}}{2}\sin\frac{k_z}{2}.
%\end{aligned}
%\end{equation}
We also include the $A_{1g}$, $B_{2g}$, and $E_g$ SOC terms,
\begin{equation}
\begin{aligned}
H_\text{SOC}&=
%& i\lambda \sum_{k_x,i_y} \sum_{abl} \epsilon_{abl} c_{C,k_x,i_y,\sigma}^{a\dagger}c_{C,k_x,i_y,\sigma'}^b \hat{\sigma}_{\sigma\sigma'}^l \\
%&+i\sum_{k_x,i_y}\lambda_{k_x}^{B_{2g}}\sigma_{\sigma\sigma'}^y
%c_{C,k_x,i_y,\sigma}^{xz\dagger}c_{C,k_x,i_y,\sigma'}^{xy} \\
%&-i\sum_{k_x,i_y}\lambda_{k_x}^{B_{2g}}\sigma_{\sigma\sigma'}^x
%c_{C,k_x,i_y,\sigma}^{yz\dagger}c_{C,k_x,i_y,\sigma'}^{xy} \\ 
%&+i\sum_{k_x,i_y}\lambda_{k_x}^{E_{g},x}\sigma_{\sigma\sigma'}^z
%c_{C,k_x,i_y,\sigma}^{xz\dagger}c_{C,k_x,i_y,\sigma'}^{xy} \\
%&-i\sum_{k_x,i_y}\lambda_{k_x}^{E_g,y}\sigma_{\sigma\sigma'}^z
%c_{C,k_x,i_y,\sigma}^{yz\dagger}c_{C,k_x,i_y,\sigma'}^{xy},
 i\lambda \sum_{\textbf{k},i_y} \sum_{abl} \epsilon_{abl} c_{C,\textbf{k},i_y,\sigma}^{a\dagger}c_{C,\textbf{k},i_y,\sigma'}^b \hat{\sigma}_{\sigma\sigma'}^l \\
&+i\sum_{\textbf{k},i_y}\lambda^{B_{2g}}(k_x)\sigma_{\sigma\sigma'}^y
(c_{C,\textbf{k},i_y,\sigma}^{xz\dagger}c_{C,\textbf{k},i_y+1,\sigma'}^{xy}  \\
&\quad-c_{C,\textbf{k},i_y,\sigma}^{xz\dagger}c_{C,\textbf{k},i_y-1,\sigma'}^{xy}) \\
&-i\sum_{\textbf{k},i_y}\lambda^{B_{2g}}(k_x)\sigma_{\sigma\sigma'}^x
(c_{C,\textbf{k},i_y,\sigma}^{yz\dagger}c_{C,\textbf{k},i_y+1,\sigma'}^{xy} \\ 
&\quad-c_{C,\textbf{k},i_y,\sigma}^{yz\dagger}c_{C,\textbf{k},i_y-1,\sigma'}^{xy}) \\
&+i\sum_{\textbf{k},i_y}\lambda^{E_{g},x}(\textbf{k})\sigma_{\sigma\sigma'}^z
(c_{C,\textbf{k},i_y,\sigma}^{xz\dagger}c_{C,\textbf{k},i_y+\frac{1}{2},\sigma'}^{xy} \\
&\quad+c_{C,\textbf{k},i_y,\sigma}^{xz\dagger}c_{C,\textbf{k},i_y-\frac{1}{2},\sigma'}^{xy}) \\
&-i\sum_{\textbf{k},i_y}\lambda^{E_g,y}(\textbf{k})\sigma_{\sigma\sigma'}^z
(c_{C,\textbf{k},i_y,\sigma}^{yz\dagger}c_{C,\textbf{k},i_y+\frac{1}{2},\sigma'}^{xy} \\
&\quad-c_{C,\textbf{k},i_y,\sigma}^{yz\dagger}c_{C,\textbf{k},i_y-\frac{1}{2},\sigma'}^{xy}),
\end{aligned}
\end{equation}
where $\epsilon_{abl}$ is the completely antisymmetric tensor,
the momentum dependent $B_{2g}$ SOC, $\lambda^{B_{2g}}(k_x) = 2i\lambda_{B_{2g}}\sin k_x$, and
the momentum dependent $E_{g}$ SOCs, 
$\lambda^{E_{g},x}(\textbf{k}) = 4\lambda_{E_{g}}\sin \frac{k_{x}}{2} \sin \frac{k_z}{2}$ and
$\lambda^{E_{g},y}(\textbf{k}) = 4i\lambda_{E_{g}}\cos \frac{k_{x}}{2} \sin \frac{k_z}{2}$.
Finally, the hopping between regions $B$ and $C$ has the form,
\begin{equation}
        \xi_{B/C}^a(k_x) = t_{B/C}^a -2t_{B/C}^{'a}\cos k_x.
\end{equation}

\begin{table}[htp]
\begin{center}
\begin{tabularx}{\columnwidth}{>{\centering\arraybackslash}X >{\centering\arraybackslash}X >{\centering\arraybackslash}X >{\centering\arraybackslash}X >{\centering\arraybackslash}X}
    \hline
    \hline
    $t_1$ & $t_2$ & $t_3$ & $t_4$ & $t_5$ \\
    0.51 & 0.06 & 0.5 & 0.18 & 0.02 \\
    \hline
    $t_6$ & $t_7$ & $t_z^{1D}$ & $t_z^{xy}$ & $\mu_\text{1D}$\\
    -0.01 & 0.009 & -0.025 & 0.002 & 0.52 \\
    \hline
    $\mu_{xy}$ & $\lambda$ & $\lambda_{B_{2g}}$ & $\lambda_{E_g}$ & $t$  \\
    0.63 & 0.1 & $-0.02$ & $-0.0025$ & 0.5 \\                           
    \hline
    $t'$ & $\mu_1$ & $\mu_2$ & $t_{B/C}^{yz}$& $t_{B/C}^{xz}$  \\ 
    0.2 & 0.18 & $-10$ & 0.05 & 0.5\\
    \hline
    $t_{B/C}^{xy}$ & $t_{B/C}^{'yz}$ &$t_{B/C}^{'xz}$&$t_{B/C}^{'xy}$&\\ 
    0.5  & 0.02 & 0.2 &0.2&\\
    \hline
    \hline
\end{tabularx}
\end{center}
\caption{
Tight-binding parameters used in all of the calculations presented
in this work.
The parameters for the SRO region were obtained from 
the DFT calculations presented in Ref.~\onlinecite{Clepkens2021PRB}, but
limited to nearest and next-nearest neighbor terms.
}\label{tbparams}
\end{table}

\begin{figure}%[H]
\includegraphics{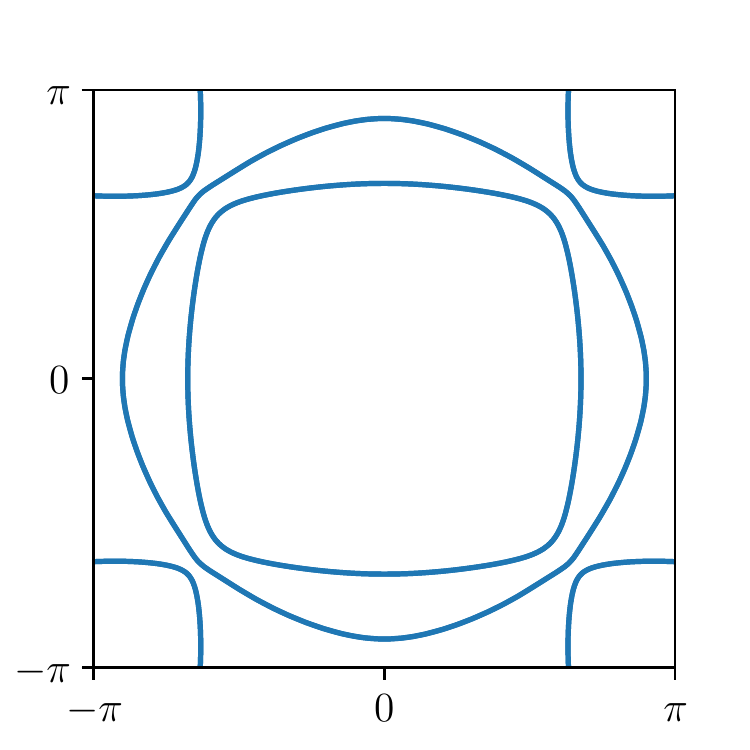}
\caption{
        Fermi surface of SRO shown in the $k_z=0$ plane, 
        for the tight-binding parameters presented in 
        Table I.
}\label{FS}
\end{figure}

The SRO tight-binding parameters used in all of the calculations presented here
were obtained from Ref.~\onlinecite{Clepkens2021PRB}, and are listed in
Table \ref{tbparams}.
The SOC values used here are %different than those found in 
chosen to be similar to the values
%Ref.~\onlinecite{Clepkens2021PRB}, but are similar to the values 
at which higher-angular momentum pairings were found to exist.
The Fermi surface obtained using the listed tight-binding parameters 
is shown in Fig.~\ref{FS} for the $k_z=0$ plane.

The irreducible representation of each of the pairing states discussed
in this work in terms of the interorbital parameters shown in Eq.~(\ref{Delta})
can be found in Ref.~\onlinecite{Clepkens2021PRB}.  
In all calculations presented here we fix the 
magnitude of the $|\vec\Delta_{xz/xy}|=|\vec\Delta_{yz/xy}^l|=1\times10^{-3}$
for consistency. In the $s$-wave case which also features finite
$|\Delta_{yz/xz}^z|$, this value is set to $5\times10^{-4}$, all in units 
of $2t=1$.
Note that while the interorbital terms are chosen to be equal for all 
interorbital pairing states, this does not mean that the gap size is equal,
as the gap size and nodal structure are determined by various parameters such
as the SOC and dispersion terms.

\section{Two Orbital Analysis}\label{2orbap}

Insights into the mechanism by which the odd-parity behavior 
appears can be gained by cosidering a two-orbital model which can be solved
analytically. 
The two-orbital Hamiltonian is presented in Eq.~(\ref{2orb}) shown in Section \ref{iop}.
Following the procedure presented in Ref.~\onlinecite{Clepkens2021PRB}, 
this Hamiltonian is tranformed into the band basis using,
\begin{equation}
    \begin{bmatrix}
        c_{\textbf{k},\sigma}^a \\
        c_{\textbf{k},\sigma}^b
    \end{bmatrix}
    =
    \begin{bmatrix}
            f_{\textbf{k}\sigma} & -g_{\textbf{k}\sigma} \\
            g_{\textbf{k}\sigma} & f_{\textbf{k}\sigma}^*
    \end{bmatrix}
    \begin{bmatrix}
        c_{\textbf{k},\sigma}^{\alpha} \\
        c_{\textbf{k},\sigma}^{\beta}
\end{bmatrix}.
\end{equation}
%where $\eta_\sigma = \pm1$ for $\sigma=\upparrow,\downarrow$.
The bands are denoted by $\alpha$ and $\beta$, %$s$ is the pseudospin
index, and the transformation coefficients are,
\begin{equation}
\begin{aligned}
f_{\textbf{k}\sigma}&=\frac{-t(\textbf{k})-i(\eta_\sigma\lambda(\textbf{k}))-\alpha(\textbf{k})}{\sqrt{t(\textbf{k})^2+
(\eta_\sigma\lambda(\textbf{k})-\alpha(\textbf{k}))^2}}
\sqrt{\frac{1}{2}\bigg(1+\frac{\xi^-(\textbf{k})}{
%\sqrt{\xi^-(k)^2+4[t(k)^2+
%(\eta_\sigma\lambda(k)+\alpha(k))^2]}
E_\sigma'(\textbf{k})
}\bigg)},\\
g_{\textbf{k}\sigma}&=
-\sqrt{\frac{1}{2}\bigg(1-\frac{\xi^-(\textbf{k})}{
%\sqrt{\xi^-(k)^2+4[t(k)^2+
%(\eta_\sigma\lambda(k)+\alpha(k))^2]}
E_\sigma'(\textbf{k})
}\bigg)}.
\end{aligned}
\end{equation}
Here, $\eta_\sigma=\pm1$ for $\sigma=\uparrow,\downarrow$ and 
$E_\sigma'(\textbf{k})=\sqrt{\xi^-(\textbf{k})^2+4(t(\textbf{k})^2+(\eta_\sigma\lambda(\textbf{k})^2
-\alpha(\textbf{k})^2))}$ where the energy eigenvalues are
$E_\sigma^\pm(\textbf{k})=\frac{1}{2}(\xi^+(\textbf{k})\pm E_\sigma'(\textbf{k}))$.

In the orbital basis, the pairing Hamiltonian is written,
\begin{equation}
\begin{aligned}
H_\text{SC}&=\Delta_{z}(c_{-\textbf{k},\downarrow}^bc_{\textbf{k},\uparrow}^a
-c_{-\textbf{k},\downarrow}^ac_{\textbf{k},\uparrow}^b
+c_{-\textbf{k},\uparrow}^bc_{\textbf{k},\downarrow}^a
-c_{-\textbf{k},\uparrow}^ac_{\textbf{k},\downarrow}^b).
\end{aligned}
\end{equation}
This is transformed into the band basis using the same transformation
described above.  Since we are only interested in the intraband pairing
we neglect pairing terms between the $\alpha$ and $\beta$ bands.
The pairing in the band basis is simplified using 
$f_{-\textbf{k}\downarrow}^*=f_{\textbf{k}\uparrow}$ and $g_{-\textbf{k}\downarrow}=g_{\textbf{k}\uparrow}$,
and appears as,
\begin{equation}
\begin{aligned}
H_\text{pair}&=i\Delta_{a/b}^z\{\text{Im}[g_{\textbf{k}\uparrow}f_{\textbf{k}\uparrow}+g_{-\textbf{k}\uparrow}f_{-\textbf{k}\uparrow}]
(\hat\Delta_0^\alpha(\textbf{k})-\hat\Delta_0^\beta(\textbf{k})) \\
&+\text{Im}[g_{\textbf{k}\uparrow}f_{\textbf{k}\uparrow}-g_{-\textbf{k}\uparrow}f_{-\textbf{k}\uparrow}]
(\hat\Delta_z^\alpha(\textbf{k})-\hat\Delta_z^\beta(\textbf{k}))\}.
\end{aligned}
\end{equation}
Since $f_{\textbf{k}\sigma}$ and $g_{\textbf{k}\sigma}$ are neither even nor odd
in the presence of $\alpha(\textbf{k})$, we perform a Taylor expansion 
of $\frac{1}{E_\uparrow'(\textbf{k})}$ for small $\alpha(\textbf{k})$.
\begin{equation}
\begin{aligned}
&\text{Im}[f_{\textbf{k}\uparrow}g_{\textbf{k}\uparrow}]=
-\frac{\lambda(\textbf{k})-\alpha(\textbf{k})}{2E_\uparrow'(\textbf{k})} \\
&=\frac{1}{2}(\lambda(\textbf{k})-\alpha(\textbf{k}))\bigg\{\frac{1}{\sqrt{
\xi^-(\textbf{k})^2+4(t(\textbf{k})^2+\lambda(\textbf{k})^2)}} \\
&+\frac{4\lambda(\textbf{k})\alpha(\textbf{k})}{
(\xi^-(\textbf{k})^2+4(t(\textbf{k})^2+\lambda(\textbf{k})^2))^\frac{3}{2}}+\mathcal{O}(\alpha(\textbf{k})^2)\bigg\}.
%&\text{Im}[f_{\textbf{k}\uparrow}g_{\textbf{k}\uparrow}]=
%-\frac{\lambda(\textbf{k})-\alpha(\textbf{k})}{2E_\uparrow'(\textbf{k})} \\
%&=\frac{1}{2}(\lambda(\textbf{k})-\alpha(\textbf{k}))\bigg\{\frac{1}{\sqrt{
%\xi^-(\textbf{k})^2+4(t(\textbf{k})^2+\lambda(\textbf{k})^2)}} 
%+\frac{4\lambda(\textbf{k})\alpha(\textbf{k})}{
%(\xi^-(\textbf{k})^2+4(t(\textbf{k})^2+\lambda(\textbf{k})^2))^\frac{3}{2}}+\mathcal{O}(\alpha(\textbf{k})^2)\bigg\}.
\end{aligned}
\end{equation}
%With the $\alpha(\textbf{k})$ dependence only in the numerator, 
Now, writing the
even-parity pseudospin-singlet and odd-parity pseudospin-triplet 
terms are straightforward, and are shown in Eqs.~(\ref{sgap}) and
(\ref{tgap}), respectively.
Eqs.~(\ref{intras}) and (\ref{intrat}) are obtained by instead assuming 
that the finite expectation value exists only in the intraorbital
spin-singlet pairing channel.  The same Taylor expansion is performed
for the coeffiecients $|f_{\textbf{k}\uparrow}|^2$ or $|g_{\textbf{k}\uparrow}|^2$, however
the lack of $\alpha(\textbf{k})$ in the numerator of these terms leads to odd-parity
contributions only arising from higher order terms.

%\newpage 

\end{document}